\documentclass[preprint2]{aastex}

\usepackage{hyperref}
\usepackage{amsmath,mathtools}
\usepackage{graphicx}
\usepackage{natbib}
\usepackage[usenames,dvipsnames]{color}
\usepackage{xspace}

\newcommand{\ud}[1]{\,\mathrm{d}#1}
\newcommand{\br}[1]{\left({#1}\right)}
\renewcommand{\sq}[1]{\left[{#1}\right]}
\newcommand{\abs}[1]{\left|#1\right|}

\newcommand{\mycitep}[1]{{\citep{#1}}}
\newcommand{\mycitet}[1]{{\citet{#1}}}

\newcommand{\secref}[1]{{Sect.\,\ref{#1}}}

\newcommand{\figref}[1]{{Fig.\,\ref{#1}}}
\renewcommand{\eqref}[1]{{Eq.\,\ref{#1}}}
\newcommand{\donotremove}[1]{}

\newcommand{\ml}{local mass-to-light ratio\xspace}

\begin{document}

\title{Velocity dispersion as a factor modifying the distribution of mass in disk-like galaxies -- an example of galaxy \mbox{UGC 6446}}

\date{Dec 7, 2015}

\shorttitle{Velocity dispersion in disk-like galaxies}
\shortauthors{Sz. Sikora, {\L}. Bratek, J. Ja{\l}ocha, M.
Kutschera} 

\author{
Szymon Sikora}
\affil{Astronomical
Observatory, Jagellonian University, Orla 171, PL-30244
Krak{\'o}w, Poland}

\author{
{\L}ukasz Bratek${}^1$}
\affil{Institute of Nuclear Physics, Polish Academy of Sciences,
Radzikowskego 152, PL-31342 Krak\'{o}w, Poland}
\email{Lukasz.Bratek@ifj.edu.pl}

\author{
Joanna Ja{\l}ocha}
\affil{Institute of Nuclear Physics, Polish Academy of Sciences,
Radzikowskego 152, PL-31342 Krak\'{o}w, Poland}

\author{
Marek Kutschera}
\affil{Institute of
Physics, Jagellonian University,  Reymonta 4, PL-30059 Krak{\'o}w,
Poland}

\begin{abstract}
Within the disk model framework used to approximately describe flattened galaxies, we develop an iterative method of determining column mass density from rotation curve supplemented with isotropic velocity dispersion profile. This generalizes our previous iterative method to the case when the velocity dispersion becomes important. We show on the example of \mbox{UGC 6446} galaxy, that taking the velocity dispersion into account results in some observational signatures in the behavior of the \ml. 
Along with galactic magnetic fields, this is another factor allowing to substantially reduce the \ml at galactic outskirts. Taking the velocity dispersion into account  may also have some consequences for the division of mass distribution between various mass components in modeling rotation curves.
\end{abstract} 


\maketitle

\section{\label{sec:intro}Introduction}
Dynamical models of spiral galaxies often adopt several constraining assumptions to simplify the relations between the disk mass density, the observed brightness, and the velocity dispersion. For example, based on the  \mycitet{1981A&A....95..105V} results, \mycitet{2008AJ....136.2782L} relates the stellar vertical velocity dispersion $\sigma_{\ast,z}$ and the stellar disk mass density $\Sigma_\ast$, by assuming the stellar disk to be isothermal in the vertical direction, with exponential radial and vertical profiles and the width-scale independent of the radial variable. Similar constraints are adopted by \mycitet{2012ApJ...751..123W} and \mycitet{2013A&A...557A.131M} who additionally assume a constant mass-to-light ratio as if the mass density distribution were to follow the brightness profile. The attractiveness of such assumptions lies in reducing complicated dynamical models to only several analytical formulas, leaving aside the demanding and nontrivial task of determining the mass distribution from the velocity field alone. In this context, a model based on the minimum number of constraining assumptions would give the opportunity to describe more general situations and provide a tool for assessing the relevance of assumptions made by simpler models.

Except for particular symmetry, the local mass distribution in flattened galaxies is known to be related to the fragment of rotation law (available observationally) in a non-unique way. For this reason, in the approximation of disk model, \mycitet{2008ApJ...679..373J} proposed an iterative method of reconstructing the column mass density from the rotation curve and complementary data (such as the observations of the distribution of hydrogen).  In contrast to parametric models customarily considered in modeling the rotation curves,  in this approach, neither the functional form of the density distribution is a priori assumed nor the mass-to-light ratio plays the role of a constant parameter, and once the brightness profile have been measured the \ml becomes the model prediction. 

The aim of this work is to generalize our iterative  method to situations with the velocity dispersion profile being isotropic and  given as a function of the radius: $\sigma(R)$. Although the velocity dispersion  observations are difficult, there is a hope such observations will be available in the nearest future. For example, the Disk Mass Survey provides data on the line of sight velocity dispersion 
at particular galactocentric distances for a sample of 30 galaxies \mycitep{2010ApJ...716..198B}.

The structure of this paper is the following. In \secref{sec:model} 
we present the theoretical framework of our extended iterative method. Next, in \secref{sec:example}, 
we apply this method to a particular galaxy which we analyzed in the past. Then conclusions follow.

\section{\label{sec:model}Determining the mass distribution by means of iterations.}
We consider an axi-symmetric and infinitesimally thin disk in the plane $z=0$ with polar coordinates $(R,\phi)$ in the framework of Jeans theory. The disk surface mass density $\Sigma$, the velocity components $v_R$, $v_\phi$ and the gravitational potential $\Phi$ are related through the continuity equation, the equations of hydrodynamical form and the Poisson equation. The term on the right hand side of the Jeans equation
involving the velocity dispersion $\sigma^2$, can be regarded as the source of an additional force per unit mass: $\vec{f}=\frac{1}{\Sigma}\nabla(\Sigma\,\sigma^2)$. When the functional form of the mass density $\Sigma(R)$ is not given a priori,  the determination of the mass distribution from the velocity field becomes nontrivial. As the starting point for further analysis, in the section below we begin with a simpler case when $\vec{f}=0$.

\subsection{The case with the vanishing velocity dispersion}
In the approximation of circular orbits ($v_R=0$), the surface mass density $\Sigma$ and the azimuthal velocity component $v$ can be related one to another through integral transformations. Recently, \mycitet{2015MNRAS.451.4018B} gave a particularly simple and symmetric form of such transformations in explicit form:
\begin{equation}\label{eqn:transformV}
v^2(R)=\int\limits_{0}^{\infty}w(x)\,\mu(R\,x)\,\ud x\,,
\end{equation}
\begin{equation}\label{eqn:transformS}
\mu(R)=\int\limits_{0}^{\infty}w(x)\,v^2(R\,x^{-1})\,\ud x\,,
\end{equation}
where $\mu(R)=2\pi G\,R\,\Sigma(R)$ is a ring density. The integral kernel $w(x)$ involves the complete elliptic integrals of the first and second kind, $K$ and $E$,\footnote{$K(k)=\int\limits_{0}^{\pi/2}\frac{\ud{\varphi}}{\sqrt{1-k^2\sin^2{\varphi}}}$,\ \ \   $E(k)=\int\limits_{0}^{\pi/2}\ud{\varphi}{\sqrt{1-k^2\sin^2{\varphi}}}$
}  and is singular at $x=1$:
$$w(x)=\frac{1}{\pi}\left(\frac{K[k(x)]}{1+x}+\frac{E[k(x)]}{1-x}\right)\,,\quad k(x)=\frac{2\sqrt{x}}{1+x}\,. $$Accordingly, the integration in \eqref{eqn:transformV} and \eqref{eqn:transformS} should be understood in the Cauchy principal value sense. 

For a rotation law $v(R)$ known in advance, $\Sigma(R)$ could be derived  directly from \eqref{eqn:transformS}. Unfortunately, galactic rotation curves are measured within a finite range of radii. Cutting off the integration in \eqref{eqn:transformS} would result in an uncontrolled error in $\Sigma(R)$. 

In the literature are known integral expressions equivalent to \eqref{eqn:transformS} on infinite domain $0<x<\infty$, for example \mycitep{1963ApJ...138..385T}. One should be aware of that these various integral forms when cut off give different results and the corresponding cutoff errors are also different. A more detailed discussion is given by \mycitet{bib:bratek_MNRAS}. 

The uncontrolled cutoff errors can be reduced under some circumstances, in particular, when some data complementary to the rotation curve $v_c(R)$ are available (such as the hydrogen column density observations $\Sigma_H(R)$). At the galactic outskirts, where the density of stars is low enough, $\Sigma_H(R)$ is the main contribution to the total disk density $\Sigma$. Then, the iterative scheme proposed by \mycitet{2008ApJ...679..373J} or equivalent -- which   
based on the $v_c(R)$ and $\Sigma_H(R)$ observations and the integral transforms \eqref{eqn:transformV}, \eqref{eqn:transformS} -- leads to a global density profile $\Sigma_0(R)$ (that continuously overlaps with $\Sigma_H(R)$ in the outer regions) together with the corresponding global velocity profile $v_0(R)$ (consistent with the rotation curve $v_c(R)$ out to the cutoff radius). 

\medskip

In the subsequent section, we assume that the global profiles $\Sigma_0(R)$ and $v_0(R)$ have been already derived with the help of any method analogous to that in \mycitep{2008ApJ...679..373J}. Then, we show how to modify the $\Sigma_0(R)$ further so as the isotropic velocity dispersion $\sigma(R)$ could be accounted for.

\subsection{The case with a non-vanishing velocity dispersion}\label{sec:procedure}
For axisymmetric planar motion on circular orbits, $v_R=0$, the radial part of Jeans equations can be reduced to:
\begin{equation}\label{eqn:Euler_radial}
v_\phi^2=R\frac{\ud{}}{\ud{R}}\br{\Phi}+\Sigma^{-1}\,
R\frac{\ud}{\ud R}\br{\Sigma\,\sigma^2}\,. 
\end{equation}
In the reduction of the axi-symmetric second moment  Jeans equations to this single form, we assumed isotropic dispersion in the disk plane and that the system is reflection-symmetric with respect to that plane  (the azimuthal component of these equations and   the continuity equation are identically satisfied under the assumed symmetries).
Under these conditions, assuming the bulk density to be of a step function form: 
 $2\Sigma(R)/H$ for $\abs{z}<H/2$ and $0$ for 
$\abs{z}>H/2$, and on integrating the $z$ component of the second moment Jeans equation in the direction normal to the disk plane (with a relation $\partial_z\Phi=2\pi G\,\Sigma$ for an infinitesimally thin disk taken into account), we would end up with $$\sigma_z^2=\pi G\,H\,\Sigma$$ as an estimation of the additional vertical dispersion $\sigma_z$ needed to support a thin disk of a small width $H$ against its gravitational compression in the vertical direction. Out to a geometric factor of the order of unity (depended on the vertical structure of the thin disk), this is a well known formula often used to infer the column mass density $\Sigma$ on the basis of the vertical dispersion $\sigma_z$ and a fixed, constant disk scale height $H/2$.  Our study differs at that point from the usual approach, because we will derive the disk mass density $\Sigma(R)$ in a way independent of the disk's vertical structure (then $\sigma_z$ becomes a prediction within a given model of the vertical mass distribution). 

We find it convenient to rewrite the square of the azimuthal velocity as a sum of a gravitational part $v_\Phi^2$ and a contribution from a velocity disperssion $v_\sigma^2$: $v_\phi^2=v_\Phi^2+v_\sigma^2$. With a formal identification 
\begin{equation}\label{eqn:v_Phi}
v_\Phi^2\equiv R\frac{\ud\Phi}{\ud R},
\end{equation}
\begin{equation}\label{eqn:v_sigma}
v_\sigma^2\equiv \frac{R}{\Sigma}\frac{\ud}{\ud R}\left(\Sigma\,\sigma^2\right),
\end{equation}
the equation \eqref{eqn:Euler_radial} is automatically satisfied ($v_\sigma^2$ is only a formal symbol which may attain also negative values).
The mass density $\Sigma(R)$ is present in both of the above expressions. 
Now, with $v_\Phi^2$ substituted in place of $v^2$ in the integral \eqref{eqn:transformV}, we would be able to simultaneously account both for \eqref{eqn:v_Phi} and the Poisson equation. Accordingly, for a given $\Sigma(R)$, we would obtain $v_\Phi^2$ by evaluating the integral \eqref{eqn:transformV}, then we would make use of \eqref{eqn:v_sigma} to get the resulting $v_\sigma^2$. 

However, more interesting in the context of this paper is the  opposite task aimed at reconstructing the unknown profile $\Sigma(R)$ in
such a way that, for a given dispersion profile $\sigma(R)$, the velocity $v_\phi$ would overlap with the rotation curve. Below, we present
an iterative scheme realizing this idea.

As the starting point of the iterative procedure, we specify a velocity dispersion profile in a form $\sigma(R)=\sigma_0\,f(R)$, with $f(R)$ being  a given function kept unaltered during the iterations. The $\sigma_0$ is a variable parameter. We start with a small $\sigma_0$ value, and then we subsequently increase it by a small amount $\delta\sigma_0$ in each iteration step,  until a desired value of $\sigma_0$ is reached.

The initial conditions are set by a small value of $\sigma_0$ and a $\Sigma_0(R)$ profile obtained iteratively at the vanishing velocity dispersion. With these initial data, further iteration steps (with their own  fixed $\sigma_0$) consist of three stages: (\emph{i}) we calculate the velocity squared $v_{\sigma\,(1)}^2$ from \eqref{eqn:v_sigma} with $\Sigma_0$ substituted in place of $\Sigma$, (\emph{ii}) we insert the resulting $v_{\sigma\,(1)}^2$ into the integral transform \eqref{eqn:transformS} to obtain a correction to the density profile $\delta\Sigma_{1}$ and the resulting first corrected density $\Sigma_1(R)=\Sigma_0(R)-\delta\Sigma_1$, (\emph{iii}) we calculate the velocity squared $v_{\sigma\,(2)}^2$ from \eqref{eqn:v_sigma} with $\Sigma_1$ substituted in place of $\Sigma$. It is right to expect that $v_{\sigma\,(2)}^2$ will differ from $v_{\sigma\,(1)}^2$. To minimize this difference, we insert $v_{\sigma\,(2)}^2$ into  \eqref{eqn:transformS}, which gives rise to a further correction $\delta\Sigma_{2}$ and, hence, a second corrected density $\Sigma_2(R)=\Sigma_0(R)-\delta\Sigma_2$. By going on with repeating the steps (\emph{i}-\emph{ii}) until the difference $v_{\sigma\,(k)}^2-v_{\sigma\,(k-1)}^2$ converges to within an acceptable limit, we obtain a $k$-th corrected profile $\Sigma_k(R)=\Sigma_0(R)-\delta\Sigma_k$ compatible with the velocity dispersion with a small fixed $\sigma_0$. It is easy to convince oneself,  that by applying the integral transform \eqref{eqn:transformV} to both sides of the equation $\Sigma_k(R)=\Sigma_0(R)-\delta\Sigma_k$, one would end up with $v_\Phi^2=v_0^2-v_\sigma^2$ by construction.  
Once $\Sigma(R)$ is obtained for a small $\sigma_0$, we may repeat the steps (\emph{i})-(\emph{iii}) starting with the previously obtained profile $\Sigma(R)$ and a slightly changed amplitude $\sigma_0+\delta\sigma_0$. This way, we consecutively obtain a sequence of density profiles $\Sigma(R)$ corresponding to a sequence of $\sigma_0$ values. 

To see in action the procedure just described,  we apply it below to galaxy \mbox{UGC 6446}.

\section{\label{sec:example}Application to galaxy \mbox{UGC 6446}}
We studied galaxy \mbox{UGC 6446} already in  \mycitep{2010MNRAS.406.2805J} where we gave some arguments suggesting why this galaxy should be regarded as a flattened object rather than dominated by a spherically symmetric component. On the other hand, the \ml of this galaxy grew rapidly at the edge of the stellar disk, suggesting a domination of unseen masses. Therefore, we find \mbox{UGC 6446} to be a good candidate for studying possible sensitivity of the model predictions to additional factors  -- such as the velocity dispersion -- modifying the mass density-rotation dependence and the resulting mass-to-light ratio behavior.

\subsection{The case with the vanishing velocity dispersion and a comment on  smoothing}
\begin{figure}[h]
   \centering
      \includegraphics[width=0.4\textwidth,trim={1cm 0cm -1cm -0cm}]{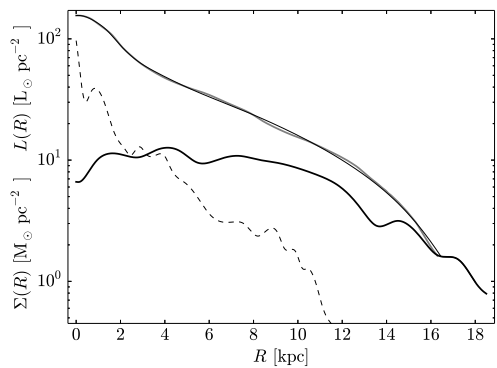}
      \caption{\label{fig:density_profile}The \mbox{UGC 6446} disk mass density $\Sigma_0(R)$ derived from the rotation curve by the iterative method without the velocity dispersion. The \emph{thin solid line} represents a smoothed cubic spline interpolation of $\Sigma_0(R)$, and it is compared with the usual cubic spline shown with the \emph{grey line}. The \emph{thick solid line} is the gas column mass density $(4/3)\Sigma_H(R)$. In the same plot, the \mycitet{1996AJ....112.2471T} B-band luminosity profile is shown with the \emph{dashed line}.}
\end{figure}

In \figref{fig:density_profile} we present a disk mass density $\Sigma_0(R)$ obtained for \mbox{UGC 6446} using a method akin to that in \mycitep{2008ApJ...679..373J}, by assuming total gas density to be that of the observed hydrogen column density $\Sigma_H(R)$ times a factor $4/3$ (to account for the helium abundance) and the 
rotation curve by \mycitet{2001A&A...370..765V}. A closer look at the luminosity distribution $L(R)$ (a dashed line in \figref{fig:density_profile}) explains why the \ml blows up at $R\approx12\,\mathrm{kpc}$. We argued \mycitep{2010MNRAS.406.2805J}  that a small fraction of unseen baryonic matter (e.g. consisting of brown dwarfs) could account for such a behavior of the \ml in the outer galaxy. In the present work we try to find out to what extent taking the velocity dispersion into account could change this picture.

It is physically clear, that when some density fluctuations appear on small scales, similar fluctuations may appear in the velocity dispersion. Below, we restrict ourselves to modeling the velocity dispersion $\sigma(R)$ by a relatively simple function, effectively neglecting fluctuations of $\sigma(R)$ on smaller scales. It is then reasonable to neglect small fluctuations in the density $\Sigma(R)$  by performing appropriate smoothing. To this end, each time a $\Sigma(R)$ is found with the help of the integral \eqref{eqn:transformS}, we tabularize the integration results at a discrete set of radii $R$. With the help of smoothed cubic spline interpolation method applied to these data, we obtain a smooth $\Sigma(R)$ profile. Although the difference between the smoothed cubic spline lines and the ordinary cubic spline lines is quite small, as seen in \figref{fig:density_profile}, this is not so for the derivatives $\Sigma'(R)$. This effect becomes important when the equation \eqref{eqn:v_sigma} is concerned.

\begin{figure}[h]
   \centering
      \includegraphics[width=0.4\textwidth,trim={1cm 0cm -1cm -0cm}]{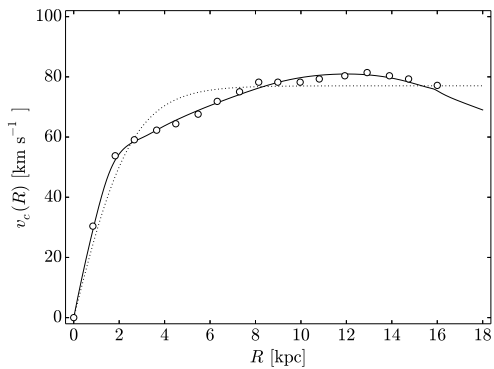}
      \caption{\label{fig:rotation} The \emph{points} represent the rotation curve of \mbox{UGC 6446} measured by \mycitet{2001A&A...370..765V}. The \emph{solid line} is the rotation curve $v_c(R)$ corresponding to the smoothed cubic spline of $\Sigma_0(R)$. The \emph{dotted line} is the best fit of the step-like profile described in the text.}
\end{figure}

The rotation curve measurements are plotted in \figref{fig:rotation}. The solid line shows the rotation velocity $v_\phi(R)$ calculated with the help of the integral \eqref{eqn:transformV} applied to the smoothed cubic spline interpolation of $\Sigma_0(R)$. The smoothing of the density profile is transferred onto the calculated rotation curve. Nevertheless, the smoothed rotation curve $v_c=v_\phi(R)$ obtained this way agrees with the original rotation data well enough. For comparison, with the dotted line in \figref{fig:rotation} we show  the best fit of the step profile $v(R)=V_{arot}\,\mathrm{tanh}(R/r_s)$. This profile was used by the Disk Mass Survey \mycitep{2013A&A...557A.130M} for a large number of other galaxies. This will help us to estimate in \secref{sec:observations} the order of magnitude of the velocity dispersion expected for \mbox{UGC 6446}.  

\subsection{A simple model of the velocity dispersion}\label{sec:step}
\begin{figure}[h]
   \centering
      \includegraphics[width=0.4\textwidth,trim={1cm 0cm -1cm -0cm}]{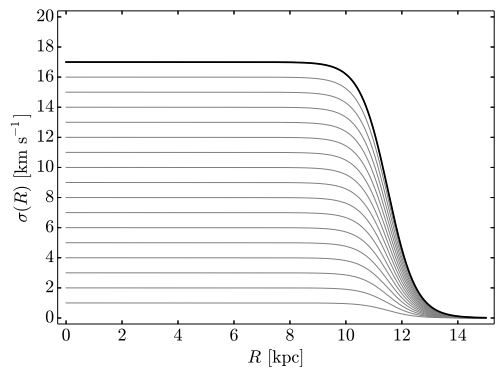}
      \caption{\label{fig:step_profile}The sequence of the velocity dispersion \emph{step-like profiles} described by Eqn \ref{eqn:dispersion_profile1} with parameters $a=1.0$ and $b=11.5$.} 
\end{figure}

The simplest choice for the velocity dispersion would be a uniform profile $\sigma=const$. But for a stationary system with circular streaming motion it is reasonable to assume that $\sigma\to 0$ at the edge of the stellar disk, since otherwise there would be a nonvanishing flux of stars at the boundary. This is why we prefer to consider the following step-like profile:
\begin{equation}\label{eqn:dispersion_profile1}
\sigma(R)=\frac{\sigma_0}{2}
\br{1-\tanh\sq{a\,(R-b)}},
\end{equation}
The arbitrary slope parameter $a$ determines the derivative $\sigma'(R)$ which should not be too high at large radii. The stellar disk's edge is determined by a radius $R_{cut}$ beyond which the mass density becomes that of gas: $\Sigma(R>R_{cut})=(4/3)\,\Sigma_H(R)$. For a fixed $a$ the parameter $b$ is determined from the requirement that $\sigma(R_{cut})$ has a given value close to zero. An example sequence of such profiles is shown in \figref{fig:step_profile} for various $\sigma_0$ and fixed $a$, $b$.

\begin{figure}[h]
   \centering
      \includegraphics[width=0.4\textwidth,trim={1cm 0cm -1cm -0cm}]{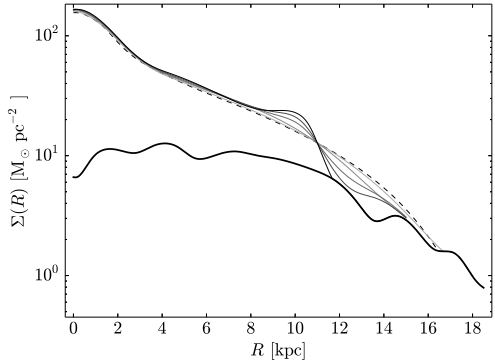}
      \caption{\label{fig:step_density}\emph{Solid lines} - the sequence of the disk density profiles $\Sigma(R)$ corresponding to the step-like velocity dispersion with parameters $\sigma_0=7, 12, 14, 16, 17\,\mathrm{km/s}$, respectively. The profiles are denoted on the gray scale from light gray for $\sigma_0=7\,\mathrm{km/s}$ to dark gray for $\sigma_0=17\,\mathrm{km/s}$. For comparison, the \emph{dashed line} is the density $\Sigma(R)$ without the velocity dispersion and the \emph{black thick line} represents the gas distribution.}
\end{figure}

\begin{figure}[h]
   \centering
      \includegraphics[width=0.4\textwidth,trim={1cm 0cm -1cm -0cm}]{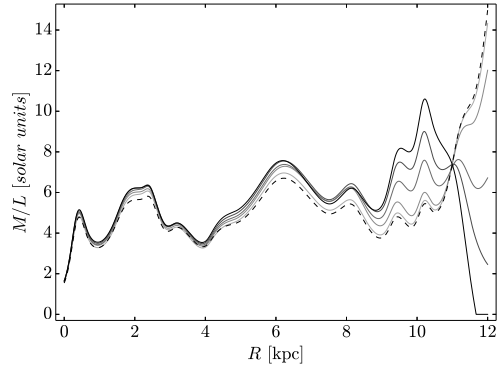}
      \caption{\label{fig:MtoL1}The \ml, shown for a sequence of density profiles $\Sigma(R)$ corresponding to a step-like velocity dispersion with parameters $\sigma_0=7, 12, 14, 16, 17\,\mathrm{km/s}$, respectively. As in \figref{fig:step_density} these lines are ordered on the gray scale, starting from light gray for $\sigma_0=7\,\mathrm{km/s}$ to dark gray for $\sigma_0=17\,\mathrm{km/s}$. The \emph{dashed line} is the \ml for the case without velocity dispersion.}
\end{figure}

By applying the iterations of \secref{sec:model} to a sequence of  velocity dispersion profiles \eqref{eqn:dispersion_profile1}, we obtain a sequence of corrected density profiles, each corresponding to a particular $\sigma_0$ value. We assumed an iteration step $\delta\sigma=1\,\mathrm{km/s}$. For clarity, in \figref{fig:step_density} the resulting  density profiles are  
presented  only for selected $\sigma_0$ values.
As can be seen, the density changes substantially at a distance $R\approx11\,\mathrm{kpc}$, thus it will have an important impact on the mass-to-light  ratio only in a neighborhood of the stellar disk's boundary.  \figref{fig:MtoL1} shows the \ml for the same sequence of $\Sigma(R)$ profiles. It can be seen, that 
by introducing the velocity dispersion to the dynamical model, 
the behavior of the \ml for \mbox{UGC 6446}  can be changed substantially at the stellar disk's edge, from that blowing rapidly up to that diminishing. Although the adopted $\sigma$ profile is very simple, the velocity dispersion gradient strongly influences the \ml in regions of low  stellar density. 

\begin{figure}[h]
   \centering
      \includegraphics[width=0.4\textwidth,trim={1cm 0cm -1cm -0cm}]{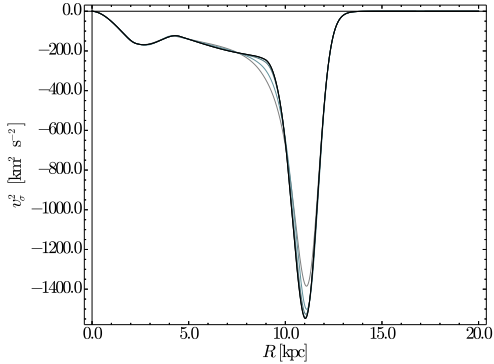}
      \caption{\label{fig:iteration}The sequence of $v_{\sigma\,(k)}^2$ lines converging to the black line in four iterations steps (\emph{i})-(\emph{iii}) described in \secref{sec:procedure}.}
\end{figure}

\begin{figure}[h]
   \centering
      \includegraphics[width=0.4\textwidth,trim={1cm 0cm -1cm -0cm}]{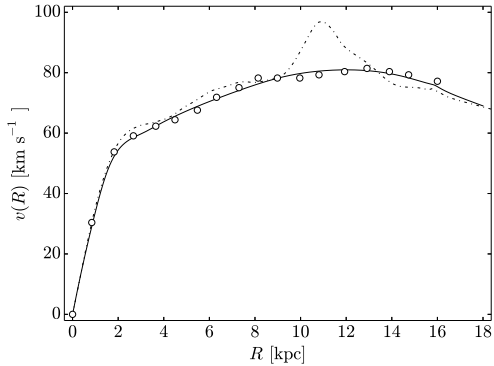}
      \caption{\label{fig:modelA_vPhi}The gravitational component of  rotation, $v_\Phi(R)$, (\emph{dashed-dotted line}) calculated for the step-like velocity dispersion profile with $\sigma_0=17\,\mathrm{km/s}$, compared with the rotation curve (\emph{solid line}) and the data points.}
\end{figure}

In order to convince ourselves that our procedure works well, we iteratively applied steps  (\emph{i})-(\emph{iii}) to find a corrected density $\Sigma_k(R)$ for an example value $\sigma_0=16\,\mathrm{km/s}$. In \figref{fig:iteration} we show a sequence $v_{\sigma\,(k)}^2$ converging quickly after four iterations only. Although the difference $v_{\sigma\,(k)}^2-v_{\sigma\,(k-1)}^2$ appears not to tend to zero for greater $k$ but to a small value (probably due to poor numerical resolution), the difference between $\sqrt{v_{\Phi\,(k)}^2+v_{\sigma\,(k)}^2}$ and $v_c$ remains lower than $\approx1\,\mathrm{km/s}$, which is enough for our purposes.
The $v_\sigma^2$ term is negative, which means that the gravitational contribution $v_\Phi$ is greater than that expected based on the rotation curve $v_c$ only. This is intuitively clear as the action of the velocity dispersion term is similar in effect to a hydrodynamical pressure that must be balanced in a stationary case by increased gravitational force. 
In view of \eqref{eqn:transformV}, the gravitational component $v_\Phi$ is the velocity for which the centrifugal force would equal the increased gravitational force. This is why we expect $v_\Phi>v_c$ and this agrees with what is seen in \figref{fig:modelA_vPhi}. The only region where $v_\Phi$ is a bit lower than $v_c$ is $R>12\,\mathrm{kpc}$. This is because we consider here the velocity dispersion of the stellar disk only, while the position $R_{cut}$ (where the density distribution is glued with the gas density) acquired a shift from the initial value $R_{cut}=14\,\mathrm{kpc}$ (the case without dispersion) to $R_{cut}=12\,\mathrm{kpc}$ of the final density distribution. 

In the next section, we will estimate the expected velocity dispersion in the \mbox{UGC 6446} galaxy on the basis of the available observations.

\subsection{Likely  $\sigma$ values for \mbox{UGC 6446}. }\label{sec:observations}
In comparing model predictions with measurements, we should keep in mind that the only direct observable at our disposal is the velocity dispersion along the line of sight $\sigma_{LOS}$. Splitting the velocity dispersion into radial and vertical components, is always affected by certain model-dependent assumptions. As the simplest solution, in this paper we assume isotropic velocity dispersion, in which case $\sigma=\sigma_{LOS}$. The advantage of this assumption lies in making the presentation of our method more transparent by avoiding unnecessary complications, but the presented procedures could be generalized also to nontrivial forms of the velocity dispersion tensor. 

We want to emphasize the importance of the Disk Mass Survey observations \mycitep{2013A&A...557A.130M} offering precise data on $\sigma_{LOS}$ as a function of radius for 30 galaxies from the UGC catalog. Unfortunately, \mbox{UGC 6446} is not included in the PPak sample. Nevertheless, we may try to estimate the likely value of the velocity dispersion based on these data. \mycitet{2013A&A...557A.130M} model rotation curves by a step-like profile:
\begin{equation}
v_c(R)=V_{arot}\,\mathrm{tanh}(R/r_s)\,,
\end{equation}
with parameters $V_{arot}$ and $r_s$ (\figref{fig:rotation} shows the best fit we obtained for \mbox{UGC 6446}, with $V_{arot}=77\,\mathrm{km/s}$ and $r_s=2.57\,\mathrm{kpc}$), and they assume the stellar $\sigma_{LOS}$ to obey an exponential decrease law:
\begin{equation}\label{eqn:exp_dispersion}
\sigma_{LOS}(R)=\sigma_{LOS,0}\,\exp(-R/h_{\sigma,LOS})\,.
\end{equation}
Following \mycitet{1981A&A....95..105V}, they also found -- that
for a mass density with exponential profile (both in the radial and vertical direction), a constant mass-to-light ratio and constant stellar velocity ellipsoid -- the radial scale parameter $h_R$ of the mass density and the velocity dispersion scale parameter $h_{\sigma,LOS}$ 
 are related by  $h_{\sigma,LOS}=2\,h_R$. 
\begin{figure}[h]
   \centering
      \includegraphics[width=0.4\textwidth,trim={1cm 0cm -1cm -0cm}]{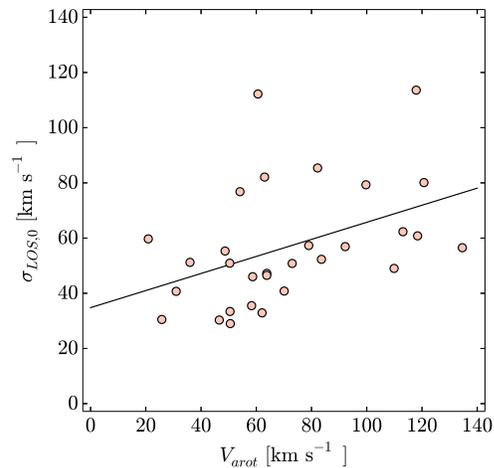}
      \caption{\label{fig:DiskMass}The velocity dispersion parameter $\sigma_{LOS,0}$ and the rotation speed $V_{arot}$ for the PPak sample of the Disk Mass Survey. The \emph{solid line} is a linear fit.}
\end{figure}
In \figref{fig:DiskMass} we show the $\sigma_{LOS,0}$ and $V_{arot}$
parameters for the 30 galaxies from the PPak sample. A linear regression $\sigma_{LOS,0}=a\,V_{arot}+b$ to these data with the best fit values $a=0.30$ and $b=34.8\,\mathrm{km/s}$, gives for \mbox{UGC 6446} an estimate of $\sigma_{LOS,0}=58.6\,\mathrm{km/s}$. With the radial scale parameter $h_R=1.87\,\mathrm{kpc}$ for \mbox{UGC 6446}   \mycitep{2001MNRAS.325.1017V}, this leads us 
to $h_{\sigma,LOS}=3.74\,\mathrm{kpc}$, under the conditions described above. With a dashed line in \figref{fig:3models_dispersion}, we show the resulting exponential dispersion profile \eqref{eqn:exp_dispersion}. We stress that this is only a rough estimate which we use as a reference profile for the three model profiles we present in the next section. It specifies the order of magnitude and the likely shape, helpful in replacing the step profile of \secref{sec:step} by those  better mimicking the velocity dispersion behavior in real galaxy.
\begin{figure}[h]
   \centering
      \includegraphics[width=0.4\textwidth,trim={1cm 0cm -1cm -0cm}]{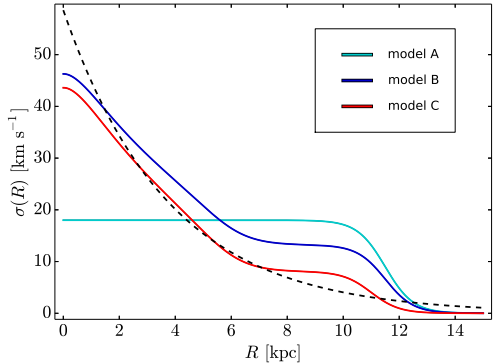}
      \caption{\label{fig:3models_dispersion}The three toy models of the velocity dispersion profile for \mbox{UGC 6446} described in detail in the text. For comparison, the \emph{dashed line} represents the exponential profile \eqref{eqn:exp_dispersion} with parameters $\sigma_{LOS,0}=58.6\,\mathrm{km/s}$ and $h_{\sigma,LOS}=3.74\,\mathrm{kpc}$ determined in \secref{sec:observations}.}
\end{figure}

\subsection{More involved dispersion models}
In this section, in addition to the velocity dispersion profile \eqref{eqn:dispersion_profile1} with $\sigma_0=17\,\mathrm{km/s}$ (which we refer to as model \textsf{A}), we consider two other profiles of the form:
\begin{equation}\label{eqn:modelBC}
\sigma(R)=\sigma_0\,\left(A(R)\,\exp(-R/h_{\sigma,LOS})+B(R)\right), 
\end{equation}
with $A(R)$ and $B(R)$ being some parametric functions. Their 
form is arbitrary to the extent that $\sigma(R)$ should be a smooth function such that: (\emph{i}) its slope in the inner galaxy part is similar to the exponential reference profile and more flattened in the center, (\emph{ii}) in the outer galaxy part the $\sigma(R)$ behaves like a step profile. To guarantee the convergence of the iterative method, we intentionally flattened the $\sigma(R)$ out at the galaxy center (when the radial derivative of $\sigma(R)$ is too high at $R=0$, the iterative procedure becomes divergent). We don't have to worry about this because the disk model does not properly approximate the more spherical  central galactic part. When the bulge is modeled separately, this problem does not occur. Figure 12 in \mycitep{2013A&A...557A.130M} shows that the velocity dispersion in the outer part is higher than shown by the exponential fit, therefore the step-like behavior of the profile at the galaxy edge is justified.   

\begin{figure}[h]
   \centering
      \includegraphics[width=0.4\textwidth,trim={1cm 0cm -1cm -0cm}]{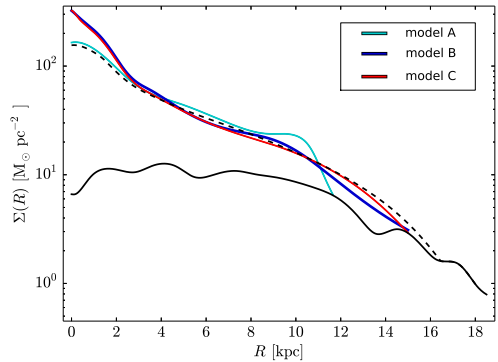}
      \caption{\label{fig:3models_density}The disk mass density $\Sigma(R)$ reconstructed iteratively for velocity dispersion models \textsf{A, B} and \textsf{C}, respectively. The \emph{dashed line} is the density distribution in the case without velocity dispersion and the \emph{thick black line} is the gas density.}
\end{figure}

\begin{figure}[h]
   \centering
      \includegraphics[width=0.4\textwidth,trim={1cm 0cm -1cm -0cm}]{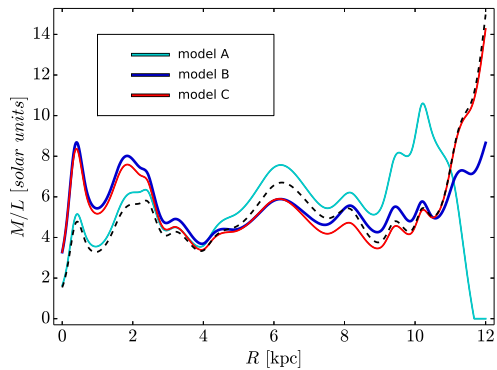}
      \caption{\label{fig:3models_MtoL}The \ml as a function of radius corresponding to the density distribution for models \textsf{A, B} and \textsf{C}, respectively. The \emph{dashed line} is the \ml for a mass density distribution without velocity dispersion.}
\end{figure}

We show in \figref{fig:3models_dispersion} two profiles of the form \eqref{eqn:modelBC} (called model \textsf{B} and model \textsf{C}) together with a step-like profile of \secref{sec:step} (called model \textsf{A}). For comparison, the reference exponential profile is also shown.  Corresponding to these models, in \figref{fig:3models_density} we show  the disk density distributions (obtained iteratively as described in \secref{sec:procedure}) and in \figref{fig:3models_MtoL} we show the \ml{s}.
It is evident from these figures, that depending on the velocity dispersion values in the stellar disk edge, various kinds of the \ml behavior are possible in the outer galaxy. The ratio may be left unchanged  compared to that with no velocity dispersion (model \textsf{C}), in which case the \ml diverges at $R=12\,\mathrm{kpc}$; it may converge to zero at the stellar disk edge neighborhood (model \textsf{A}); or become oscillating around a value characteristic of the galaxy as a whole (model \textsf{B}). This proves that the velocity dispersion is capable of modifying the \ml in the outer galactic regions. At the same time,  the mass of the galactic central part is higher when the
velocity dispersion of stars grows towards the galaxy center, so for  models \textsf{B} and \textsf{C} the \ml is slightly higher
in the central part $R<3\,\mathrm{kpc}$.

\begin{figure}[h]
   \centering
      \includegraphics[width=0.4\textwidth,trim={1cm 0cm -1cm -0cm}]{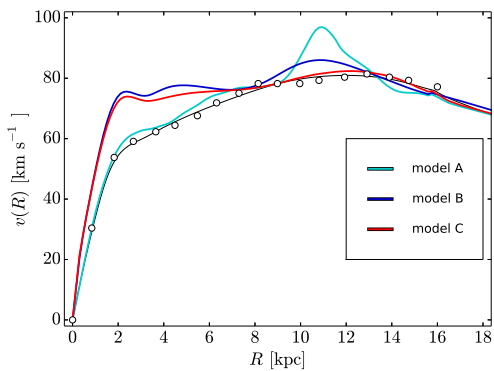}
      \caption{\label{fig:3models_vPhi}The $v_\Phi$ velocity component obtained iteratively for the \mbox{UGC 6446} galaxy with fixed model dispersion profiles \textsf{A, B} and \textsf{C} described in the text. For comparison, \emph{solid black line} is the smoothed rotation curve $v_c$ and the \emph{points} show the rotation curve measurements.}
\end{figure}
\figref{fig:3models_vPhi} shows  the gravitational part of the rotation curve $v_{\Phi}$ for the considered dispersion models. Similarly as for model \textsf{A} analyzed in \secref{sec:step}, the values attained by  $v_{\Phi}$ are higher than the corresponding values on the rotation curve $v_c$. This means that the mass distribution in the disk has been modified in comparison with the situation without dispersion, such that the resulting correction to the gravitational force counterbalances the effective pressure due to the velocity dispersion term.

\section{Conclusions}
In this work, we presented a systematic iterative method of determining  the column mass density in the disk model approximation that takes the stellar velocity dispersion into account. The input quantities consist of the rotation curve $v_c(R)$, the measured hydrogen column density $\Sigma_H(R)$ and a given velocity dispersion profile $\sigma(R)$. We assumed an axially symmetric disk and a streaming motion about concentric circular orbits. Our approach differs from other models considered in the literature in that we assume nothing  about the functional form of the mass density and its relation to the brightness profile. In our method the \ml is a variable output quantity, whereas this ratio is often considered to be a constant model parameter. This qualitative change makes our model more general and potentially useful in judging the importance of assumptions made by simpler models.

The presented method can be applied to any spiral galaxy for which $v_c(R)$, $\Sigma_H(R)$ and $\sigma(R)$ are known. We tested this method on the example of galaxy \mbox{UGC 6446}. We already studied this galaxy in the past by neglecting the velocity dispersion and found a peculiarity in the behavior of the \ml at the edge of the stellar disk. In this paper we showed  that in a region of low stellar mass density, the velocity dispersion has a substantial influence on the resulting \ml of this galaxy.  One cannot exclude the possibility that a similar effect may also occur in other galaxies.

In the context of standard modeling of spiral galaxies, which usually splits  rotation curves into several subcomponents (the bulge, disk and a dark matter halo), it is important to note 
that for non-vanishing velocity dispersion, the gravitational contribution to the rotation curve $v_\Phi$ may differ significantly from that curve,  as it is evident from \figref{fig:3models_vPhi}. This result strongly suggests, that together with the standard components of the rotation curve, also the contribution from the velocity dispersion $v_\sigma$ should be taken into account. This effect may have consequences on the division of mass distribution between various mass components, in particular, in the disk and in the external halo. Along with galactic magnetic fields \mycitep{bib:Jalocha11042012,bib:Jalocha21112012}, accounting for the non-vanishing velocity distribution is another factor allowing to substantially reduce the \ml at galactic outskirts.    

\bibliography{UGC6446}
\bibliographystyle{aa}
\end{document}